# Demonstrating levitation within a microwave cavity

Nabin K. Raut,[1, *] Jeffery Miller,[1] Raymond Chiao,[1] Jay E. Sharping,[1]

[1]Physics department, University of California, Merced, 95348 CA, USA

Levitated systems are desirable due to reduced clamping losses and reduced thermal contact[1]. These advantageous properties have been exploited in optomechanics to achieve ultra-strong coupling between the mechanical mode and the electromagnetic mode[2]. Such schemes provide an opportunity for the quantum manipulation of a macroscopic system [3]. In this letter, we report the first successful experiments with a levitated millimeter-scale neodymium magnet within a centimeter-scale superconducting aluminum coaxial quarter-wave stub cavity. The magnet levitated near the top of the stub, where the electric field is concentrated, perturbs the electric field distribution allowing for small perturbations in the magnet's position to be detected through shifts in the resonance frequency. Resonance spectra are collected via a vector network analyzer (VNA) between temperatures of 5 K and 50 mK revealing movement of the magnet inside of the cavity. Room temperature measurements and finite element calculations are done to calculate the shift in frequency for various positions of the magnet, and an experimentally measured 100 MHz upshift when transitioning into a superconducting state confirms levitation with remanences up to 140 times stronger than the critical field of the aluminum. We achieve levitation heights of 1 – 1.8 mm. We investigate the dependence of levitation height and levitation temperature on the strength of the magnet and, surprisingly, we observe that the levitation temperature and height both increase with permanent magnet strength. Our work describes a novel macroscopic mechanical system capable of sensing and transducing forces, thus allowing for the coupling of disparate classical and quantum systems.

Significant progress has been made with optical levitation[4] where a main challenge includes noise arising from photon recoil and heating[5]. A promising alternative is to use passive levitation techniques involving magnets and superconductors. In one example, a magnet levitated above a superconductor is used to achieve mechanical transduction between individual spin qubits[6]. In these passive levitation systems it is important to distinguish between Type-I vs Type-II superconductivity and whether the experiment is conducted in a zero-field cooled, or non-zero-field cooled manner. When cooling a Type-II superconductor through its superconducting transition in the presence of a non-zero magnetic field, magnetic flux is trapped by vortices within the material. The trapped flux then holds the permanent magnet in place and complicates its motion. If one cools the Type-II superconductor in the absence of magnetic fields and then introduces the permanent magnet, it is levitated, but the magnet must be inserted after cooling. In comparison, when a Type-I material makes the superconducting transition the superconductor exhibits a perfectly diamagnetic response as long as the magnetic field is less than some critical value. Supercurrents build up on the surface of the superconductor which fully screen the magnetic flux from its interior[7]. The Meissner-force is always repulsive and can be large enough to lift macroscopic objects to some equilibrium height where the net force of the interaction combined with that due to gravity is zero.

Three-dimensional superconducting radio frequency (SRF) cavities have well-defined frequency spectra whose localized electric and magnetic fields provide unique opportunities to couple to other objects such as mechanical oscillator to form a cavity optomechanical system. Because of the physical positioning of the field distributions, the losses can be minimized allowing for a large quality factor[8-10]. Depending upon the strength of the coupling, a wide variety of physical phenomena can be explored from the semi-classical Purcell effect[11] to the quantum realm (detection of gravitational waves[12, 13], entangled state generation[14-16], optomechanically induced transparency[17]) along with novel physics (dark matter detection[18], multimode entanglement[19] ). Superconducting radio-frequency cavities exhibit high power handling capability, so a strong pump field can be used for the quantum control of the mechanical element. Moreover, these cavities have been shown to be promising platforms for quantum memory and cavity QED[20-23]. Recently, there has been a growing interest in coupling a magnon with the highly concentrated RF magnetic field of the cavity[24, 25]. While coupling between mechanical oscillators and optical cavities or superconducting circuits is becoming established, we are only just beginning to see significant progress towards coupling levitated systems within bulk SRF cavities[26, 27].

Fig. 1 (a) illustrates measurements scheme of our cavity-magnet system, which sits in the mixing chamber of a dilution refrigerator at T=50 mK. The refrigerator is cooled down to approximately 50 mK (well below the zero-field critical temperature of aluminum) and vacuum pressure is maintained at $10^{-7}$ mbar. A microwave signal (~0 dBm) is coupled into the cavity to probe the quarter wave mode via a pin antenna that extends into the body of the cavity near the region with the strongest electric field. Transmission measurements ($S_{21}(\omega)$) are performed with an over-coupled cavity which results in a loaded cavity Q on the order of 2500 to ensure that the mode can be tracked on a vector network analyzer (VNA). N-type cryogenic grease is used to maximize thermal contact between the cavity and the base plate of the dilution refrigerator, and to make the cavity stable during the cooldown.

The coaxial cylindrical cavity with one end open [Fig. 1(b)] is machined out of 6061 (97.9% pure) aluminum. The outer cylinder has a radius of 7 mm and a height of 55 mm. The inner stub cylinder has radius of 2 mm and height of 5 mm giving a quarter-wave resonance of $f_0$=10 GHz. The coaxial stub cavity's resonance frequency is determined by the height of the stub, $l$, where $f_0 \alpha \frac{1}{4l}$. We use neodymium disc magnets of 1-mm diameter by 0.5-mm high, with a mass of 2.75 mg, and varying magnetic field strengths (remanence, provided by the manufacturer) of 1.22 T – 1.47 T. This corresponds to magnetic moments of 0.38-0.46 (mA)m$^2$, respectively. We include a plastic sleeve to keep the magnet on top of the stub where the frequency sensitivity is high. Without the sleeve, the magnet will fall into the gap between the stub and the wall. We expect the magnet to slide towards the edge of the stub during levitation irrespective of its initial position on the stub.

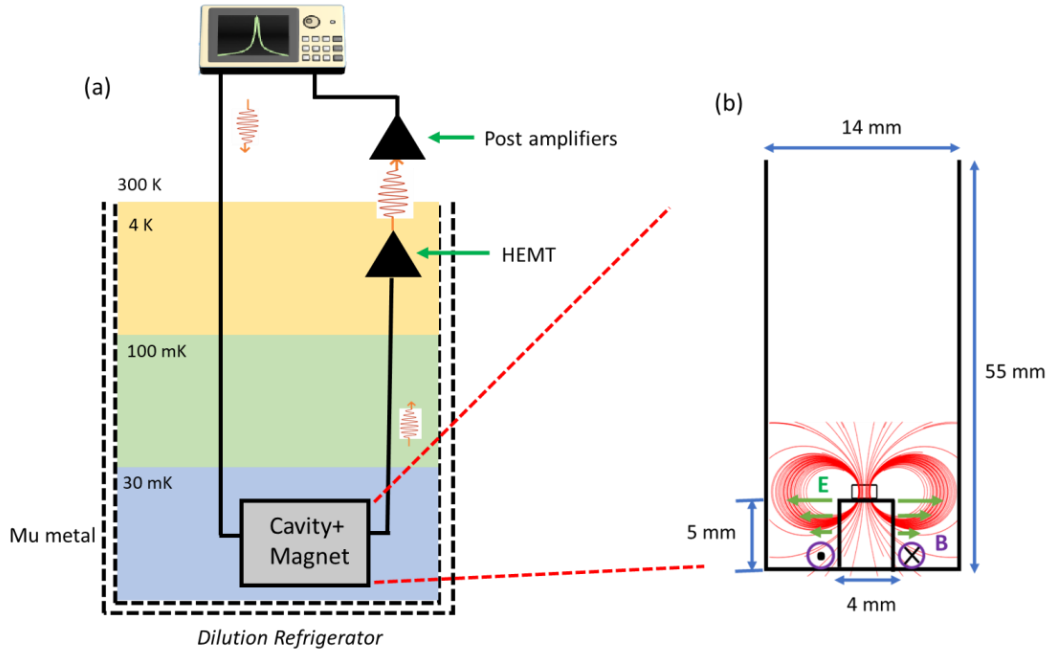

**Fig. 1| Schematic of experimental set up**. a, The cavity-magnet configuration attached to the base plate of a dilution refrigerator. b, Front view of the coaxial cavity with a magnet on the top part of the stub. The electric field highly concentrated at the rim of the stub (green arrow) decaying towards the conducting wall of the cavity.

In our experiment, any motion or levitation of the magnet results in a large frequency shift that can be observed in the VNA. Room temperature measurements and finite element calculations (COMSOL Multiphysics) help us to associate particular magnet displacements during levitation with the measured frequency shifts. The room temperature measurements are performed by manually positioning a magnet held by a capillary tube within the cavity in three dimensions. First, the capillary tube with a magnet is translated at different coordinates around and above the stub. Measurements were also made with just the capillary and no magnet to eliminate the effect of the capillary on the measurements. The frequency shift patten of these measurements compared with the simulation are displayed in Fig. 2 (a). For example, when the magnet is still in contact with the stub ($z = 0$), radial movement of the magnet towards the edge of the stub produces a frequency downshift of -50 MHz/mm. According to our calculations the largest height sensitivity is expected for a magnet positioned at the edge of the stub (1.75 mm) where the sensitivity is -400 MHz/mm for levitation heights of 0-0.1 mm. As can be seen in the overlapping traces in Fig. 2 (a), once the levitation height exceeds z=0.7 mm, the levitation sensitivity no longer depends strongly on lateral position[28]. The expected frequency shifts from these simulations are used to estimate the levitation height in the experimental data (see supplementary material for the detail levitation height calculation).

A model of how the magnet effects the effective stub height is helpful for gaining intuition about how the cavity's resonance frequency will shift in different circumstances. Figure 2 (b) illustrates the expected changes to the resonance frequency of the cavity during levitation experiments. Any perturbation within the coaxial region of the cavity changes the shape of the cavity mode and its frequency[29]. When a magnet is placed on the surface of the stub, it increases the effective height of the stub, thereby reducing the frequency of the cavity. The amount of frequency downshift depends upon the interaction of the magnet with the electric field of the cavity mode, which is concentrated toward the edges of the stub. Conversely, when the magnet is placed on the bottom of the cavity it raises floor of the cavity. This reduces the effective length of the stub which causes the resonance frequency to increase. As the magnet levitates above one of the surfaces, the resonance shifts towards that of the bare cavity (cavity without any magnet).

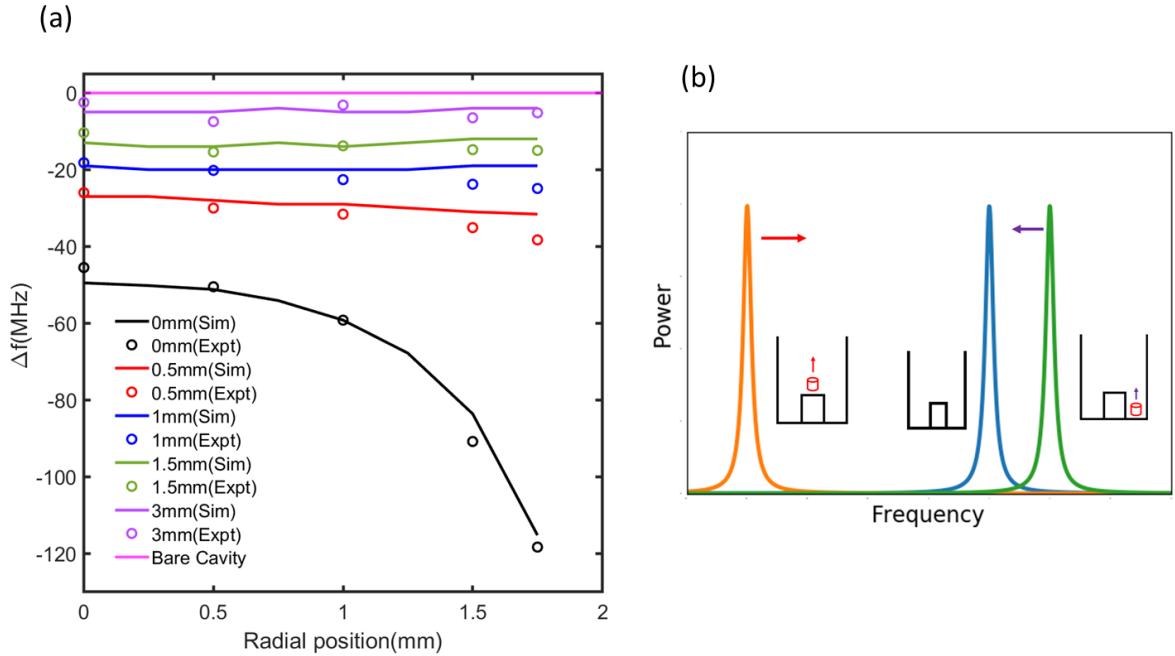

**Fig. 2| Characterization of frequency shift pattern**. a, Frequency shift away from the base resonance (horizontal pink trace) as a function of radial position for varying heights. b, Generalized frequency shift pattern with respect to the bare cavity (blue). The resonance will shift down in frequency for a magnet above the stub (gold) while an upshift is observed when the magnet is placed in the wall-stub gap (green).

The spectra collected between 1.25 K and 50 mK are compared in Fig. 3. When the bare cavity (with the plastic sleeve surrounding the stub) is cooled through the superconducting transition of aluminum at 1.2 K we observe a negligible frequency shift of a few KHz due to a change in the penetration depth. The small shifts at low temperatures, along with larger shifts (~20 MHz) due to thermal contraction as the system cools from room temperature, are reproducible across all measurements. The bare cavity's resonance at ~5K is used as the reference frequency for levitation experiments ($\Delta f=0$).

We measure the behavior of the cavity-magnet system for a sequence of four magnets having the same shape and mass but having different magnetic strengths. The magnets are placed on top of the stub but the exact positioning on the stub is undetermined. Since the stub has minor imperfections it is not perfectly symmetric and the frequency downshift due to the presence of the magnet varies depending upon the exact position of the magnet. For example, at 5K, the frequency shift due to the N35 (1.22 T) and N42 (1.32 T) magnets is ~120 MHz, but the frequency shift for the N50 (1.44 T) and N52 (1.47 T) magnets is just ~90 MHz. Multiple cooling and heating cycles are performed for each magnet (see supplementary material).

For each trace in Fig. 3, one observes four features as the temperature drops. Initially the frequency is constant while the magnet rests on the surface of the stub. Secondly there is a fluctuation region where the frequency varies from the high temperature value by ~20 MHz. Thirdly, at intermediate temperatures there is a transition region characterized by sudden transitions of >30 MHz with plateaus. And fourthly, there is a final levitation region. As an example, consider the case of the weak N35 magnet. The frequency remains constant at about -120 MHz as the temperature drops from 1.25 K to 600 mK. Between 600 mK and 300 mK the fluctuation region exhibits variations between -120 MHz and -100 MHz. Sudden large step transitions (>30 MHz) occur in the range of 300 mK down to 150 mK. Finally, the measured frequency remains steady below 100 mK. All of the magnets exhibit the four features, but magnets providing larger magnetic fields display them at higher temperatures. The frequency shifts due to the magnetic levitation are in excellent agreement with the room temperature measurements and finite element calculations (Fig. 2). Additional experiments are performed with magnets of similar strength but slightly different geometry (see supplementary material).

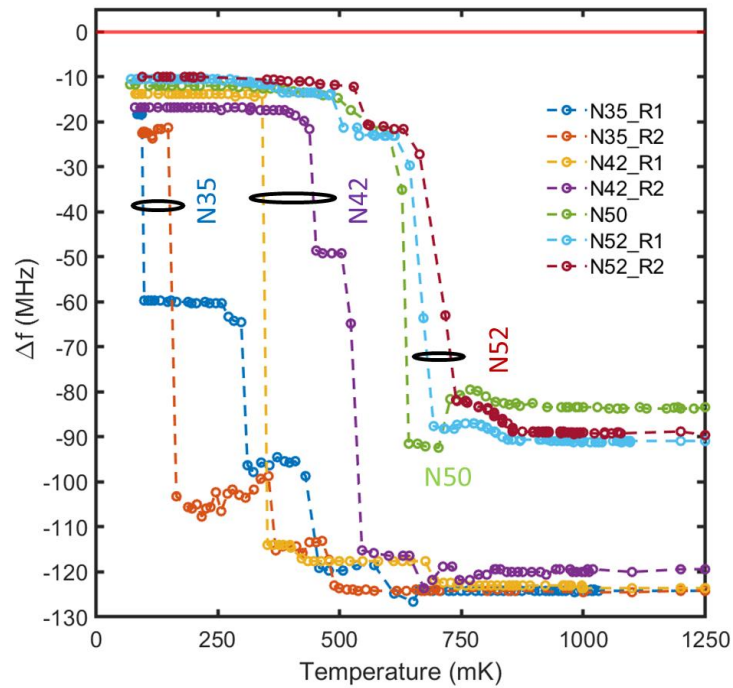

**Fig. 3| Low temperature measurements**. Change in frequency as a function of temperature for N35, N42, N50, and N52 magnets. The black ovals group together traces from the same magnet throughout multiple heating and cooling cycles.

Calculations of the potential energy for a disk magnet having a remanence of 1.47 T (N52), combined with experiments conducted without the sleeve present, predict that the minimum in the potential energy is located in the gap. In the supplementary material we introduce an extended two-loop model[30], where the cavity response is given by summing over an array of response loops. The potential energy model we develop predicts the levitation height more accurately than the widely use image method. For the N52 magnet our model overestimates levitation height only by 17 %, in contrast to 127 % overestimation by the image method (see Fig. 4).

We plot the levitation height as a function of magnet remanence in Fig. 4 where we see that magnets producing a larger magnetic field attain a larger levitation height and exhibit a higher levitation transition temperature compared with magnets producing a smaller magnetic field. This is surprising because prior to levitation, when the magnet rests directly on the surface of the aluminum, the magnetic field strength at the interface is larger than the critical field of the aluminum (10 mT at 1.2 K). This normal conducting region with a depth of 1-2 mm which varies with temperature and the strength of the magnet (see supplementary material). Levitation shows up as an abrupt transition since the levitation force depends quadratically on the magnetic moment of the magnet used. Below the levitation temperature the Meissner effect produces a force whose magnitude is sufficient to overcome the force due to gravity and allow the magnet to levitate. The uncertainty in our measurement of levitation height also decreases as the levitation force increases [31].

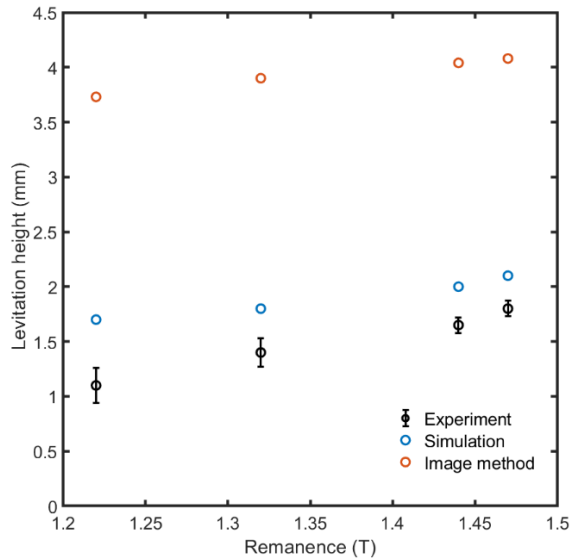

**Fig. 4| Comparison of levitation height between the image method, our model, and experimental result as a function of remanence**.

To summarize, we have characterized the levitation of mm-sized permanent magnets within a type-1 SRF cavity. We obtain levitation heights of 1-1.8 mm for commercially-available neodymium magnets. These levitation heights are in a range where the SRF cavity's resonance frequency varies with magnet position. Levitation temperatures for aluminum cavities range from 100-700 mK, which is obtainable in dilution refrigerator systems. Such electro-mechanically coupled systems, if stabilized, can be used to introduce the low-frequency mechanical motion of the magnet with other objects whose quantum states can be probed and manipulated in SRF cavities, such as magnons and transmons. In addition, the levitated high Q mechanical oscillator may be useful for precision sensing and gravitational wave detection.

We thank Prof. Chih-Chun Chien, Prof. M. Scheibner, Dr. J. Pate, Dr. L. Martinez, Dr. A. Castelli, Dr. P. Dhakal, Dr. P. Bernstein, and Prabin Parajuli for useful discussions.

*Corresponding author: nraut@ucmerced.edu

References

1. Hoang, T. M. *et al*. Torsional optomechanics of a levitated nonspherical nanoparticle. *Phys. Rev. Lett.* **117**, 123604 (2016).

2. Ahn, J. *et al*. Optically levitated nanodumbbell torsion balance and GHz nanomechanical rotor. *Phys. Rev. Lett.* **121**, 033603 (2018).

3. Aspelmeyer, M., Kippenberg, T. J. & Marquardt, F. Cavity optomechanics. *Reviews of Modern Physics* **86**, 1391 (2014).

4. Ahn, J. *et al*. Ultrasensitive torque detection with an optically levitated nanorotor. *Nature Nanotechnology* **15**, 89-93 (2020).

5. Vinante, A. *et al*. Testing collapse models with levitated nanoparticles: Detection challenge. *Physical Review A* **100**, 012119 (2019).

6. Gieseler, J. *et al*. Single-Spin Magnetomechanics with Levitated Micromagnets. *Phys. Rev. Lett.* **124**, 163604 (2020).

7. Blundell, S. J. in *Superconductivity: a very short introduction* (OUP Oxford, 2009).

8. Kuhr, S. *et al*. Ultrahigh finesse Fabry-Pérot superconducting resonator. *Appl. Phys. Lett.* **90**, 164101 (2007).

9. Reagor, M. *et al*. Quantum memory with millisecond coherence in circuit QED. *Physical Review B* **94**, 014506 (2016).

10. Ciovati, G., Dhakal, P. & Gurevich, A. Decrease of the surface resistance in superconducting niobium resonator cavities by the microwave field. *Appl. Phys. Lett.* **104**, 092601 (2014).

11. Kockum, A. F., Miranowicz, A., De Liberato, S., Savasta, S. & Nori, F. Ultrastrong coupling between light and matter. *Nature Reviews Physics* **1**, 19-40 (2019).

12. Rodgers, P. Mirror finish. *Nature Materials* **9**, S20 (2010).

13. Pace, A. F., Collett, M. J. & Walls, D. F. Quantum limits in interferometric detection of gravitational radiation. *Physical Review A* **47**, 3173 (1993).

14. Mancini, S., Giovannetti, V., Vitali, D. & Tombesi, P. Entangling macroscopic oscillators exploiting radiation pressure. *Phys. Rev. Lett.* **88**, 120401 (2002).

15. Vitali, D. *et al*. Optomechanical entanglement between a movable mirror and a cavity field. *Phys. Rev. Lett.* **98**, 030405 (2007).

16. Ockeloen-Korppi, C. F. *et al*. Stabilized entanglement of massive mechanical oscillators. *Nature* **556**, 478-482 (2018).

17. Weis, S. *et al*. Optomechanically induced transparency. *Science* **330**, 1520-1523 (2010).

18. Asztalos, S. J. *et al*. SQUID-based microwave cavity search for dark-matter axions. *Phys. Rev. Lett.* **104**, 041301 (2010).

19. Peterson, G. A. *et al*. Ultrastrong parametric coupling between a superconducting cavity and a mechanical resonator. *Phys. Rev. Lett.* **123**, 247701 (2019).

20. Ladd, T. D. *et al*. Quantum computers. *Nature* **464**, 45-53 (2010).

21. Majer, J. *et al*. Coupling superconducting qubits via a cavity bus. *Nature* **449**, 443-447 (2007).

22. Nissen, F. *et al*. Nonequilibrium dynamics of coupled qubit-cavity arrays. *Phys. Rev. Lett.* **108**, 233603 (2012).

23. Lu, Y. *et al*. Universal stabilization of a parametrically coupled qubit. *Phys. Rev. Lett.* **119**, 150502 (2017).

24. Tabuchi, Y. *et al*. Coherent coupling between a ferromagnetic magnon and a superconducting qubit. *Science* **349**, 405-408 (2015).

25. Flower, G., McAllister, B., Goryachev, M. & Tobar, M. E. Determination of niobium cavity magnetic field screening via a dispersively hybridized magnonic sensor. *Appl. Phys. Lett.* **117**, 162401 (2020).

26. Childress, L. *et al*. Cavity optomechanics in a levitated helium drop. *Physical Review A* **96**, 063842 (2017).

27. Kashkanova, A. D. *et al*. Optomechanics in superfluid helium coupled to a fiber-based cavity. *Journal of Optics* **19**, 034001 (2017).

28. Raut, N., Miller, J., Pate, J., Chiao, R. & Sharping, J. Meissner levitation of a millimeter-size neodymium magnet within a superconducting radio frequency cavity. *IEEE Trans. Appl. Supercond.* (2021).

29. Sharping, J. E. *et al*. Joints and shape imperfections in high-Q 3D SRF cavities for RF optomechanics. *Journal of applied physics* **128**, 73906 (2020).

30. Kim, K., Levi, E., Zabar, Z. & Birenbaum, L. Restoring force between two noncoaxial circular coils. *IEEE Trans. Magn.* **32**, 478-484 (1996).

31. Simon, M. D., Heflinger, L. O. & Geim, A. K. Diamagnetically stabilized magnet levitation. *American journal of physics* **69**, 702-713 (2001).